# Design of a Dual Polarized Gold-Coated Four-Channel PCF-SPR Sensor with Ultra-High Sensitivity and Broad RI Coverage

Osama Haramine Sinan[a], Ifaz Ahmed Adan[a], Mohammad Tawsif[a], Aditta Chowdhury[a,*]

[a]Department of Electrical and Electronic Engineering,
Chittagong University of Engineering and Technology (CUET),
Chattogram-4349, Bangladesh

*Corresponding author

**Abstract**

Surface plasmon resonance (SPR) sensors built on photonic crystal fibers (PCFs) have recently become a promising category of optical sensing platforms. This is primarily due to their high sensitivity, compact structural design and ability to work with a wide variety of analytes. The overall sensing performance can significantly be enhanced by carefully tailoring the fiber geometry to strengthen the coupling between the guided core mode and the plasmon mode. This study designs and thoroughly investigates a polarization-dependent SPR-based PCF sensor that delivers strong mode coupling, wide detection range and high refractive-index (RI) sensitivity using finite element method (FEM). In the proposed configuration, air holes with different diameters and a rectangular core slot are used to create birefringence along with four outer microchannels coated with a 30 nm gold film create distinct plasmonic regions that support multiple resonances. The simulated results yield maximum wavelength sensitivities of 77500 $\text{nmRIU}^{-1}$ and 92500 $\text{nmRIU}^{-1}$, along with amplitude sensitivities of 3930 $\text{RIU}^{-1}$ and 3139 $\text{RIU}^{-1}$ and figures of merit of 517 $\text{RIU}^{-1}$ and 356 $\text{RIU}^{-1}$ for the x- and y-polarizations, respectively, over an analyte RI range of 1.32–1.42. The parametric study verifies that the design is highly tolerant to fabrication deviations and the coupling behavior is primarily sensitive to the smallest air holes and the height of the core slot. Since the sensing region is external, analytes can be reused conveniently. Consequently, the proposed sensor provides a compact, highly sensitive and practically realizable platform for biochemical and chemical sensing applications.

## 1. Introduction

The demand for compact, highly sensitive and reliable sensors has grown rapidly in areas such as environmental monitoring [1], biochemical testing and medical diagnosis [2]. However, conventional electronic sensors often struggle with limitations like susceptibility to electromagnetic interference, limited sensitivity and incompatibility with miniaturized or remote systems [3]. As a result, researchers have explored alternative solutions. Recent studies indicate that optical fiber-based sensing technologies provide several benefits over electronic sensors, including compact size and immunity to electromagnetic interference. They also support long-distance signal transmission and offer stable operation in harsh or corrosive environments [4], [5].

Among the different optical fiber architectures, photonic crystal fiber (PCF) has become a reliable choice for researchers to develop advanced sensors [6]. Standard optical fibers depend on total internal reflection. In contrast, PCFs guide light through a regular pattern of air holes built into the cladding [7]. This periodic microstructure can be adjusted to modify the propagation of light. Due to this distinctive geometry, PCFs enable strong light–matter interaction. As a result, PCFs have been widely used for sensing refractive index, temperature, pressure and various chemical parameters [8], [9], [10].

To further strengthen the interaction between the guided mode and surrounding analytes, plasmonic materials such as gold, silver or graphene are commonly integrated into PCF configurations [7]. When a thin metallic film is deposited near the fiber core, the incident light can excite surface plasmon polaritons (SPPs) at the metal–dielectric boundary. The resulting surface plasmon resonance (SPR) arises when the momentum of the core-guided mode becomes phase-matched with that of the SPP mode. This leads to a sharp resonance peak in the transmission spectrum [11]. The position and amplitude of this resonance are extremely sensitive to variations in the refractive index (RI) of the adjacent analyte. This mechanism forms the underlying operating principle of PCF-SPR sensors. Such sensors have been extensively investigated for chemical, biochemical and environmental monitoring, exhibiting outstanding sensitivity and fast response characteristics [12], [13].

Despite these advantages, many existing PCF-SPR sensor architectures continue to face both practical and performance related constraints. For example, D-shaped or side-polished fiber geometries often demand intricate fabrication procedures, including high-precision polishing, metallic coating and accurate alignment. These processes increase manufacturing cost and diminish reproducibility which complicates large-scale production [14], [15]. Additionally, numerous reported designs deliver high sensitivity only within a narrow RI window. This limits their effectiveness for low-RI analytes such as gases, organic solvents, or dilute biofluids. Furthermore, some sensor structures suffer from unstable modal coupling or insufficient mechanical robustness, thereby limiting their deployment in real-world conditions. Collectively, these challenges underscore the necessity for an improved PCF-SPR configuration that can offer wide detection range, higher sensitivity and simplified fabrication while preserving strong structural integrity.

In response to these challenges, the proposed design introduces a new four-channel photonic crystal fiber-based surface plasmon resonance sensor. The designed sensor incorporates several circular microchannels arranged around the core, each functioning as an individual plasmon-supporting region. This arrangement facilitates the excitation of multiple resonance modes which allows broadening the usable refractive index detection range. Moreover, an asymmetric rectangular air hole embedded at the fiber core induces birefringence which enhances polarization discrimination and improves sensing resolution. Positioning the analyte layer on the exterior surface simplifies sample introduction, cleaning and reuse. That way the requirement for complicated analyte infiltration into internal microstructures is eliminated. Due to strong coupling between the core-guided mode and the plasmonic modes, the proposed sensor achieves high wavelength and amplitude sensitivities over a wide span of analyte refractive indices. As a result, the design not only streamlines fabrication but also strengthens practical suitability. The proposed sensor presents a promising candidate for next-generation optical sensing in chemical, biomedical and industrial applications.

## 2. Literature Review

Photonic crystal fiber–based surface plasmon resonance (PCF-SPR) sensors have become a highly active research area for detecting chemical and biological analytes because they offer extremely high sensitivity, tunable structural parameters and compact configurations. To improve sensing accuracy and widen the refractive-index (RI) detection window, numerous studies have explored different fiber geometries, plasmonic materials and coating strategies to strengthen the coupling between the guided core mode and the surface plasmon polariton (SPP) modes. Nevertheless, many reported designs continue to suffer from restricted sensing ranges, moderate sensitivities or fabrication challenges.

A dual-core PCF-SPR sensor utilizing a silver-oxide ($AgO_2$) plasmonic film for wide-range RI detection was reported by Hussain et al. [16]. Their model achieved a wavelength sensitivity of 24 300 nm/RIU and a figure of merit (FOM) of 120 $RIU^{-1}$. Although the optical response was strong, the use of silver caused chemical instability and oxidation and the single-polarization operation limited measurement accuracy. In [17], a gold-coated rectangular-core PCF sensor supporting dual-polarization detection in the 1.33–1.43 RI range was presented. The sensor attained wavelength sensitivities of 58 000 nm/RIU and 62 000 nm/RIU for x- and y-polarized modes, respectively, but its sensing window remained narrow and the birefringence was not easily tunable.

To improve chemical stability, hybrid plasmonic coatings have been explored. In [18], a graphene–silver coated PCF-SPR structure achieved an amplitude sensitivity of 418 $RIU^{-1}$ and a wavelength sensitivity of 3000 nm/RIU for high-index analytes (1.46–1.49 RIU). The graphene layer enhanced electron transfer and reduced oxidation, though performance degraded for low-index samples. Similarly, Li et al. [19] proposed a graphene–gold coated exposed-core PCF-SPR sensor for the 1.333–1.3688 RI range, where adding graphene increased sensitivity from 1900 nm/RIU to 2290 nm/RIU. However, fabrication remained difficult due to non-uniform graphene deposition.

Several other designs have sought to boost modal coupling by structural optimization. Ying et al. [20] demonstrated a gold-grating-coated dual-core D-shaped PCF-SPR architecture with average wavelength sensitivity of 994.5 nm/RIU and peak amplitude sensitivity of 181 $RIU^{-1}$ in the 1.33–1.37 RI range. Although a two-feature interrogation technique improved resolution ($2.03 \times 10^{-6}$ RIU), the side-polishing process remained complex. Liu et al. [21] later proposed an asymmetric dual-core D-shaped PCF-SPR sensor using silver, achieving 14 660 nm/RIU wavelength sensitivity and 1222 $RIU^{-1}$ amplitude sensitivity, but its fabrication and limited RI window again posed challenges.

Recent works have continued to optimize performance metrics such as resolution, FOM and analyte accessibility. For instance, in [21], an internally coated PCF-SPR sensor achieved 10 000 nm/RIU wavelength sensitivity and 1115 $RIU^{-1}$ amplitude sensitivity over the 1.35–1.40 RI range. An externally coated variant in [22] expanded the detection window to 1.18–1.36 RI with 20 000 nm/RIU wavelength sensitivity and high amplitude response. The design in [23] realized 34 000 nm/RIU wavelength sensitivity with $2.94 \times 10^{-6}$ RIU resolution, while the sensor in [24] pushed further to 53800 nm/RIU wavelength sensitivity and an impressive FOM of 105 $RIU^{-1}$, highlighting the trend toward external analyte interaction for stronger field overlap. In contrast, a narrow-range, high-FOM structure in [25] achieved 2380 $RIU^{-1}$ amplitude sensitivity but operated only between 1.29 and 1.34 RI. Another external sensing configuration in [26] provided moderate performance (5100 nm/RIU, FOM = 29) yet maintained good fabrication repeatability. The work in [27] reported a simpler geometry offering 13 000 nm/RIU wavelength sensitivity across 1.20–1.28 RI, while [28] presented a design suitable for mid-index analytes (1.36-1.392 RI) with amplitude sensitivity of 909 $RIU^{-1}$.

Although these studies demonstrate steady improvement, the trade-off between high sensitivity and broad detection range remains unresolved. Many of the above sensors operate effectively either for high-index or low-index analytes, but not both and the majority depend on complex side-polishing or metal-grating processes that hinder mass fabrication. Furthermore, silver-based coatings often suffer from oxidation, while gold-based coatings though stable require improved field coupling to reach ultra-high sensitivity.

Considering these gaps, the present work proposes an enhanced four-channel gold-coated PCF-SPR sensor that simultaneously achieves broad RI coverage and high wavelength and amplitude sensitivities with a moderate FOM. The design introduces multiple plasmonic channels for extended resonance, an asymmetric rectangular air hole to induce birefringence for polarization-resolved sensing and an outer analyte layer for easy cleaning and reusability. This structure provides a practical balance between fabrication simplicity, robustness and superior optical performance, positioning it as a promising candidate for biochemical and environmental refractive-index sensing.

## 3. Numerical design and modeling

This section explains the design, simulation and practical realization of the proposed PCF-SPR sensor. Section 3.1 describes the numerical modeling process and the main performance parameters used to evaluate the sensor. Section 3.2 presents the experimental setup that demonstrates how the sensor can be practically implemented and tested. Section 3.3 discusses the fabrication feasibility, showing that the proposed structure can be produced using available fiber manufacturing and thin-film coating techniques.

### 3.1 Numerical Modeling and Performance Parameters

The design employs a circular silica substrate with carefully patterned, multi-size air apertures to enhance interaction between the guided and plasmonic modes. The schematic of the PCF SPR model appears in Fig. 1. Three distinct air-hole sizes are introduced d = 1.35, $d_1$ = 1.20 and $d_2$ = 0.18 where each serving a specific optical purpose. Smaller holes help reduce confinement losses, whereas those with larger diameters improve the field interaction between the core and the plasmonic coating. The air-hole lattice has a pitch of $\Lambda$ = 1.8 μm and at the center, a rectangular slot with height(H) and width(W) of 0.5 and 0.15 μm respectively, is embedded to break structural symmetry and induce birefringence which enhances sensing capability [29].

Around the periphery of the silica region, four circular microchannels ($d_3$ = 1.56 μm) are included to establish multiple plasmonic layers for interacting with various analyte refractive indices. These outer cavities enable the generation of multiple resonances and thus widen the usable sensing range. The fiber is designed with polishing on all four sides; the spacing between the first ring of air holes and the polished edge is $\Lambda_2$ = 2.22 µm. The gap separating the innermost air-hole ring and the gold layer is $\Lambda_1$ = 1.6 µm. This type of geometric optimization provides efficient mode excitation and stronger surface-plasmon interaction as well.

In Fig.1, the designed sensor is illustrated. The sensor is composed of multiple layers. At its center lies a silica ($SiO_2$) region. This region guides light through total internal reflection. Surrounding this core, a thin gold coating of thickness $t_g$ = 30 nm. This coating serves as the plasmon-supporting medium. An analyte region with thickness $t_a$ = 0.9 µm is placed directly above the metal film. It enables interaction between the guided and surface plasmon modes. To minimize back reflections during simulation, a perfectly matched layer (PML) (thickness $t_p$ = 1 µm) encloses the model. It ensures that outgoing electromagnetic waves are effectively absorbed.

In this configuration, the analyte layer is positioned on the outer surface of the sensor which allows it to be cleaned and reused easily for different analytes. This design avoids the fabrication challenges often faced by D-shaped or internally coated fiber sensors that require complex surface polishing [30], [31].

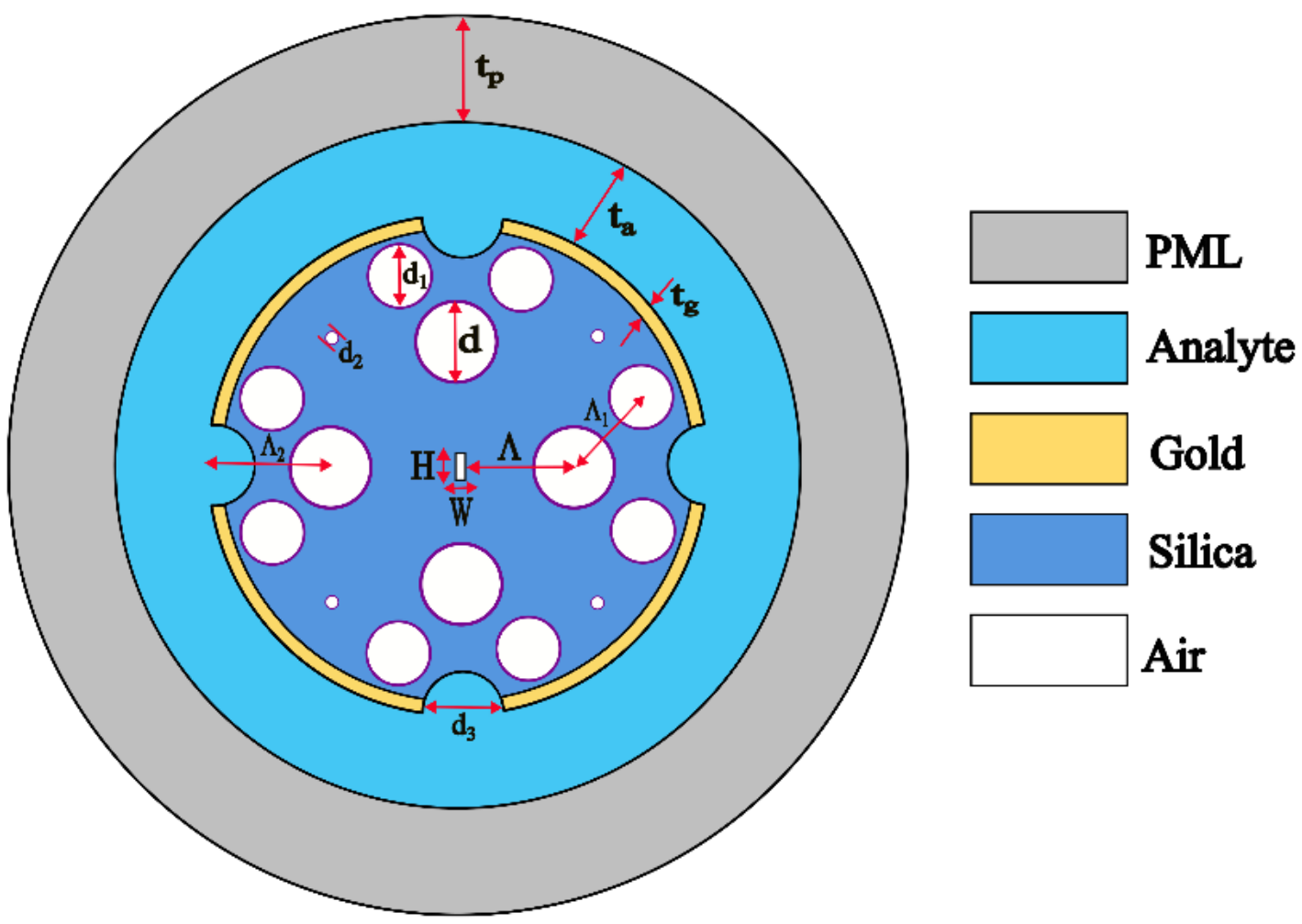

**Fig. 1:** Schematic two-dimensional layout of the proposed SPR sensor

The sensor is numerically analyzed through the finite-element method (FEM) implemented in COMSOL Multiphysics v6.2. Structural dimensions are adjusted to strengthen coupling between the guided core mode as well as the plasmonic resonance while keeping confinement loss to a minimum. Maxwell's equations are solved for each element by means of the FEM solver [32]. To eliminate back reflections at the simulation boundaries, a perfectly matched layer (PML) is applied. This layer absorbs outward-propagating electromagnetic waves as well as effectively simulates an open region. Thus it improves numerical stability and reduces computational load as well.

The wavelength-dependent refractive index of silica is obtained using the Sellmeier relation [33],

$$n = \sqrt{1 + \frac{A\lambda^2}{\lambda^2 - D} + \frac{B\lambda^2}{\lambda^2 - E} + \frac{C\lambda^2}{\lambda^2 - F}} \tag{1}$$

where $n$ denotes the RI of silica and $\lambda$ is the wavelength (µm), $A = 0.692, B = 0.408, C = 0.897, D = 0.0047, E = 0.014, F = 97.934$. The permittivity of gold, used as the plasmonic coating, follows the Drude–Lorentz formulation [34],

$$\epsilon_m = \epsilon_\infty - \frac{\omega_D^2}{\omega(\omega + j\gamma_D)} - \frac{\Delta_\epsilon \Omega_L^2}{(\omega^2 - \Omega_L^2) + j\Gamma_L \omega} \tag{2}$$

Here, $\varepsilon_\infty = 5.9673$ is the high-frequency permittivity of gold, $\Delta_\epsilon = 1.09$ is the weighting factor and $\omega$ represents the angular frequency of the guided wave. The parameters $\omega_p/2\pi = 2113.6\text{THz}$ and $\gamma_D/2\pi = 15.92\text{THz}$ correspond to the plasma and damping frequencies, respectively. Likewise, $\Omega_L/2\pi = 650.07\text{THz}$ and $\Gamma_L/2\pi = 104.86\text{THz}$ describe the Lorentz-oscillator resonance and its spectral width.

The confinement loss of the structure can later be determined by applying the imaginary component of the effective refractive index according to the following equation [35],

$$\alpha_{loss} = 8.686 \times \frac{2\pi}{\lambda} \times Im(n_{eff}) \times 10^4 \text{ dB/cm} \tag{3}$$

where $\alpha_{\text{loss}}$ represents the confinement loss, $\lambda$ represents wavelength (µm), $\text{Im}(n_{\text{eff}})$ denotes imaginary part of the complex effective refractive index.

The sensor's performance is assessed using two interrogation techniques: amplitude interrogation (AI) and wavelength interrogation (WI). For the WI approach, the sensitivity is obtained using Eq. (4) [36],

$$S_W(\lambda) = \frac{\Delta\lambda_{peak}}{\Delta n_a} \tag{4}$$

where, $\Delta\lambda_{peak}$ and $\Delta n_a$ indicates the resonance peak shifts and changes in analyte RI respectively.

In amplitude interrogation (AI), the sensor's amplitude sensitivity is obtained using the following Eq. [37],

$$S_A(\lambda) = -\frac{1}{\alpha(\lambda, n_a)} \frac{\delta\alpha(\lambda, n_a)}{\delta n_a} \tag{5}$$

where the difference observed in the confinement loss curve because of minor variation in analyte RI is indicated by $\delta\alpha(\lambda, n_a)$ and the change in analyte RI is denoted by $\delta n_a$.

Another important parameter, resolution of a sensor is calculated using the following Eq. [38],

$$R = \frac{\Delta n_a \Delta\lambda_{min}}{\Delta\lambda_{peak}} \quad (6)$$

where, R represents the sensor resolution, $\Delta n_a$ represents the variation of analyte RI, $\Delta\lambda_{min}$ defines the minimum wavelength resolution and $\Delta\lambda_{peak}$ determines the difference in resonance peak shift. To evaluate the noise immunity of the sensor and sharpness of the loss curve figure of merit is widely used performance metric. To determine figure of merit (FOM) of the sensor the following Eq. is used [39].

$$FOM = \frac{S(\lambda)}{FWHM} \quad (7)$$

Here $S(\lambda)$ is the wavelength sensitivity of the sensor and FWHM is the full width half maximum of the loss curve.

## 3.2 Experimental Setup for Practical Realization

A schematic layout for the practical realization of the designed sensor is presented in Fig. 2. In this setup, light from a broadband or supercontinuum source is injected into a single-mode fiber (SMF). The SMF is connected to the proposed PCF-SPR device through a standard splicing joint to ensure smooth optical transmission. Around the sensing region, a miniature analyte channel is formed so that the liquid sample can flow uniformly over the plasmonic surface. A micro-pump system controls the inlet (IN) and outlet (OUT) of the analyte, maintaining a constant flow rate during operation to ensure steady interaction between the analyte and the sensing layer.

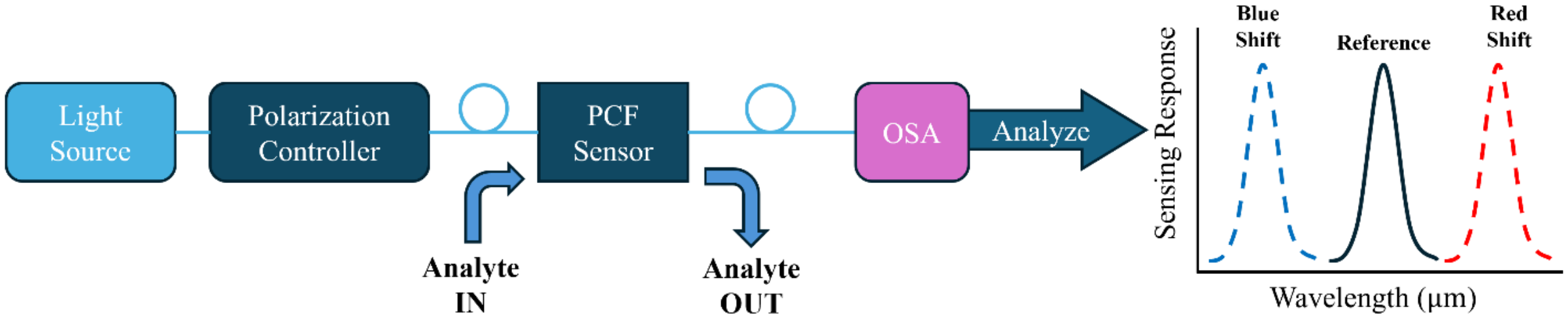

**Fig. 2:** Experimental setup of the sensor

When the analyte interacts with the plasmonic layer, the local refractive index in the vicinity of the metal surface is modified, thereby influencing the excitation of the surface plasmon polariton (SPP) mode. Thereby, whenever the refractive index (RI) changes because of the analyte, the guided light spectrum shifts. The shift can move toward a longer wavelength which is redshift or toward a shorter wavelength which is blueshift. This shift depends on how the index varies.

After the light passing through the sensing region, it is sent into a separate single mode fiber (SMF). Then the resulting spectrum is measured with an optical spectrum analyzer (OSA). Finally, the recorded data are analyzed on a computer to figure out how the sensor is responding.

## 3.3 Fabrication Feasibility

The suggested sensor is simple to fabricate and can be done with current fiber manufacturing methods. The conventional stack-and-draw technique can be used to create the variously sized circular air holes [40], [41]. Knight et al. [40] and Mahdiraji et al. [41] successfully fabricated photonic crystal fibers with circular air holes using this approach. The rectangular air hole at the core can be produced through the extrusion technique [42] or modern 3D printing technology, as demonstrated in [43], [44]. Techniques include wheel polishing, high-pressure microfluidic chemical deposition and chemical vapor deposition (CVD) can be used to apply the thin layer of gold to the outside [45]. Hence, the proposed sensor design is compatible with current fabrication capabilities.

## 4. Result and discussion

The suggested PCF SPR design is assessed via detailed finite element simulations to quantify its performance. Section 4.1 examines the modal properties, birefringence effects and polarization-dependent sensing behavior based on the evanescent field, dispersion and confinement loss characteristics. Section 4.2 investigates how variations in key structural and material parameters affect the loss spectrum and sensing response, followed by a comparison with reported sensors to highlight the improved performance and practical applicability of the proposed design.

### 4.1 Modal and Polarization-Dependent Performance Analysis

The sensing principle of the proposed PCF SPR device is governed by the evanescent field generated as guided light propagates through the fiber core. This exponentially decaying field penetrates beyond the core boundary to engage the plasmonic coating, where it stimulates plasmonic electron oscillations at the metal–dielectric interface and excites surface plasmon waves. Owing to the deliberately asymmetric fiber geometry, birefringence arises; consequently, the optical response is analyzed separately for both polarization modes.

At first, the evanescent field patterns for both polarizations are examined as shown in Fig. 3(a-d). These field distributions clearly show that the core-guided mode strongly interacts with the surface plasmon polariton (SPP) mode for each polarization. This phenomenon indicates that the plasmonic waves are being efficiently excited at the sensing interface.

Then the dispersion curves of the fundamental core mode and the plasmonic mode together with their corresponding confinement-loss characteristics are shown in Fig. 3(e-f). A distinct loss peak appears at the wavelength where the effective refractive indices (RI) of the core and SPP modes become equal. It confirms strong phase matching between them. Such sharp resonance features reflect accurate mode coupling as well as play an essential role in achieving high sensing precision.

For an analyte refractive index of $n_a = 1.41$, the resonance wavelengths are located at 1.125 μm for x-polarization and 1.175 μm for y-polarization, confirming the birefringent nature of the proposed structure. The induced birefringence facilitates selective excitation of each polarization which in turn improves the overall sensitivity and detection performance [35].

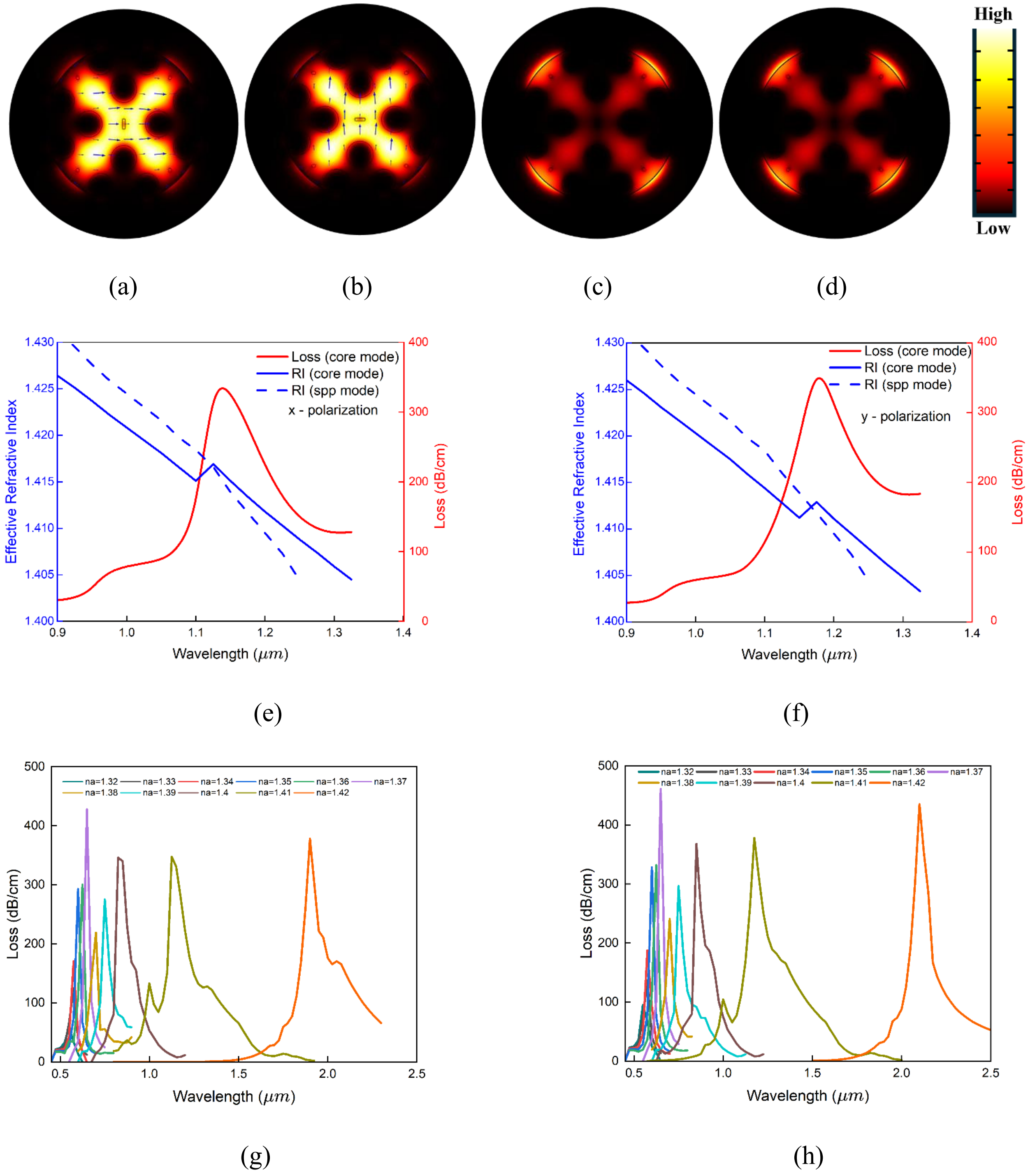

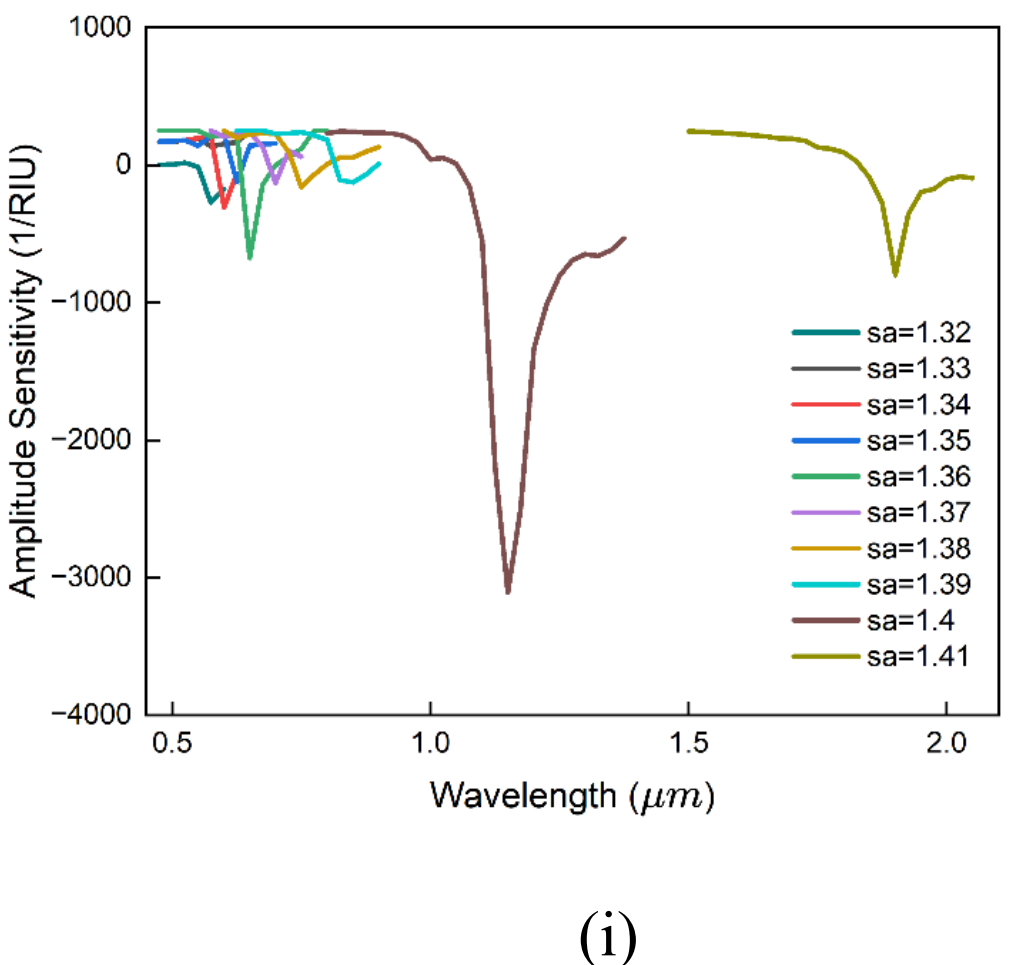

(i)

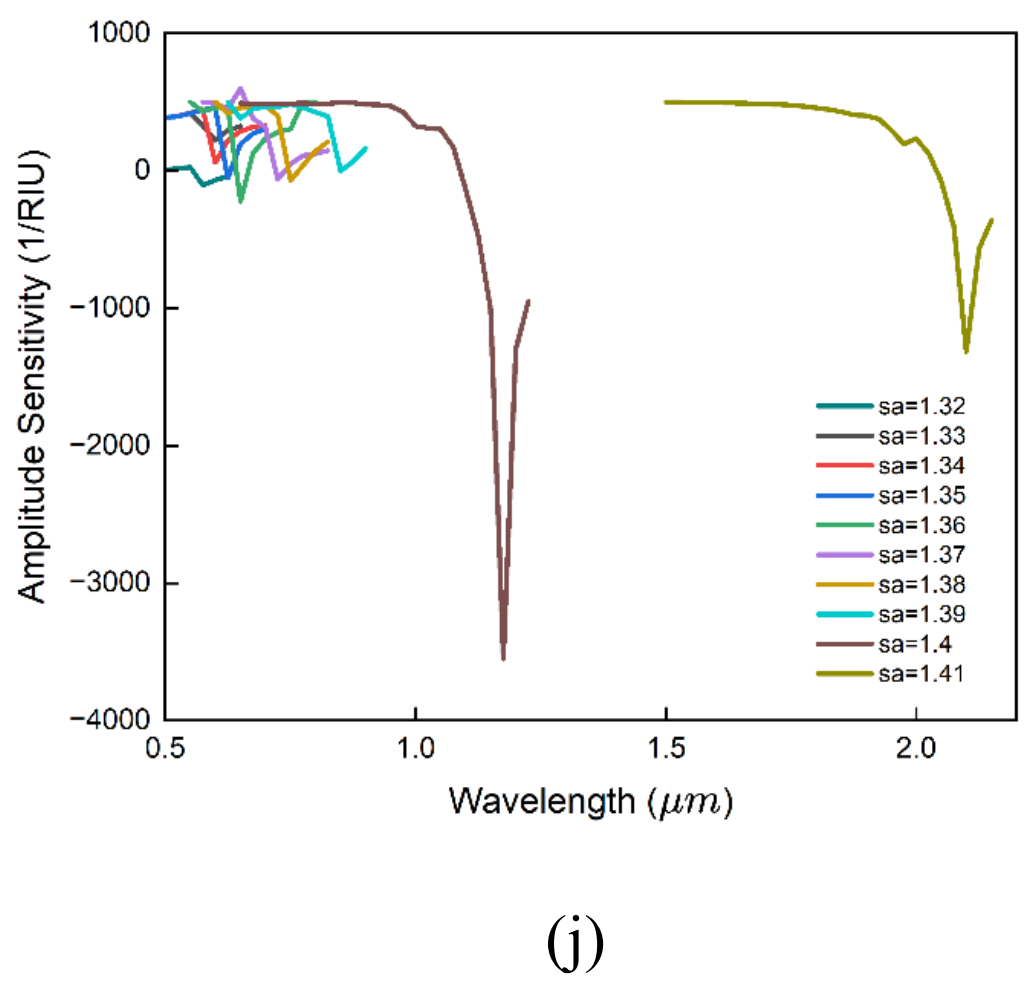

(j)

**Fig. 3:** (a–b) Electric-field patterns of the core mode; (c–d) field profiles of the SPP mode at $n_a = 1.36$; (e–f) dispersion curves and loss spectra at $n_a = 1.41$; and (g–j) confinement-loss and amplitude-sensitivity characteristics for analyte refractive indices ranging from 1.32 to 1.42 under x- and y-polarizations.

Figures 3(g–h) illustrate the confinement loss behavior for both polarization states as refractive index of the analyte varies between 1.32−1.42. A progressive redshift in resonance wavelength is observed with increasing analyte RI. This shift occurs because changes in the analyte refractive index modify the effective refractive index of the plasmonic mode, thereby shifting the phase-matching condition. The chosen RI range (1.32–1.42) encompasses most biochemical sensing regimes, where typical analyte refractive indices lie between 1.33 and 1.35 [49], [50]. A greater resonance shift indicates enhanced sensitivity, demonstrating that the proposed sensor provides strong wavelength-tuning capability. The resonance displacement of approximately 775 nm (x-pol) and 925 nm (y-pol) is recorded between $n_a = 1.41$ and 1.42 yielding highest wavelength sensitivities of 77,500 $nmRIU^{-1}$ and 92,500 $nmRIU^{-1}$. These results seem to be comparable to or surpass the performance of previously reported high-sensitivity designs [41].

To evaluate the sensor's behavior more completely, amplitude interrogation was used to determine the amplitude sensitivity for both polarization states, as shown in Fig. 3(i-j). Between these two, the x-polarized mode has sharper and deeper loss dip than the y-polarized mode which results in higher amplitude sensitivity. The obtained maximum sensitivities are 3930 $RIU^{-1}$ for x-polarization and 3139 $RIU^{-1}$ for y-polarization.

The sensor's overall detection capability was also assessed using the figure of merit (FOM). FOM is defined as the ratio of wavelength sensitivity to the full width at half maximum (FWHM) of the loss curve [36]. A higher FOM indicates better detection precision as well as more reliable signal interpretation. The FOM values corresponding to various analyte refractive indices (RI) are summarized in Table 1. According to the results as shown in the Table 1, the sensor achieves peak FOMs of 517 and 356 for the x and y polarizations, respectively. These findings show that the proposed structure outperforms several earlier plasmonic sensor designs reported in the literature [47].

Table 1 further compares the sensing behavior for both polarization states under optimized structural parameters. The x-polarized mode reaches the highest amplitude sensitivity of 3930 $RIU^{-1}$, owing to its narrower and deeper loss peak relative to the y-polarized response (Fig. 3(g-h)). The strongest amplitude sensitivity appears when the analyte refractive index is close to 1.40, then gradually decreases at higher or lower values. This behavior results from the sharper resonance near an RI range of 1.40-1.41, while the loss curve broadens for other analyte indices.

**Table 1.** Assessment of sensing characteristics using amplitude and wavelength sensitivity parameters across multiple analyte refractive indices.

| Analyte RI | Peak Wavelength (nm) | | Wavelength Sensitivity (nm/RIU) | | Amplitude Sensitivity ($RIU^{-1}$) | | FOM ($RIU^{-1}$) | |
|---|---|---|---|---|---|---|---|---|
| | x-pol | y-pol | x-pol | y-pol | x-pol | y-pol | x-pol | y-pol |
| 1.32 | 540 | 540 | 1000 | 1000 | 271 | 105 | 8 | 16 |
| 1.33 | 550 | 550 | 2500 | 2500 | 37 | 122 | 51 | 47 |
| 1.34 | 575 | 575 | 2500 | 2500 | 569 | 252 | 66 | 47 |
| 1.35 | 600 | 600 | 2500 | 2500 | 347 | 343 | 56 | 69 |
| 1.36 | 625 | 625 | 2500 | 2500 | 1010 | 478 | 72 | 64 |
| 1.37 | 650 | 650 | 5000 | 5000 | 357 | 350 | 141 | 135 |
| 1.38 | 700 | 700 | 5000 | 5000 | 394 | 357 | 89 | 83 |
| 1.39 | 750 | 750 | 7500 | 10000 | 350 | 301 | 99 | 104 |
| 1.4 | 825 | 850 | 30000 | 32500 | 3930 | 3139 | 370 | 264 |
| 1.41 | 1125 | 1175 | 77500 | 92500 | 1161 | 1351 | 517 | 356 |
| 1.42 | 1900 | 2100 | N/A | N/A | N/A | N/A | N/A | N/A |

## 4.2 Influence of Structural and Material Parameters on Sensor

The impact of structural and material parameters on confinement loss and sensing characteristics is examined by considering an analyte with an arbitrary RI of 1.36. Figure 4(a–f) illustrates how variations in geometrical dimensions influence mode coupling behavior and the resonant response.

In Fig. 4(a), changing the diameter d of central air hole from 1.33 to 1.36 μm causes a slight increase in confinement loss with negligible wavelength shift, indicating that the design preserves stable performance even under small fabrication deviations which is evidence of good tolerance. A similar trend can be seen in Fig. 4(b) for the surrounding air-hole diameter $d_1$ (1.20-1.23 μm). Only small changes in amplitude are noticed, while resonance wavelength remains essentially fixed.

In contrast, Fig. 4(c) shows that the smallest air hole $d_2$ has a stronger effect on mode coupling behavior. When $d_2$ is reduced from 0.22 μm to 0.16 μm, both the confinement loss and the resonance wavelength increases. This indicates that the interaction between the core mode and the plasmonic mode as well become stronger mainly because the evanescent field penetrates more deeply toward the metal interface.

Figures 4(d) and 4(e) illustrate how the rectangular core slot dimensions H and W effect the mode characteristics. Increasing the slot height H from 0.3 $\mu m$ to 0.6 $\mu m$ strengthens the coupling and results in a higher loss peak, while the resonance wavelength stays almost unchanged. In contrast, varying the slot width W from 0.13 $\mu m$ to 0.16 $\mu m$ shows only a small impact. This phenomenon confirms that the slot height plays a more dominant role in determining polarization behavior as well as birefringence.

In practical fabrication, the gold film thickness ($t_g$) can vary slightly during deposition. Figure 4(f) presents the influence of $t_g$ on sensor performance. A thinner layer ($t_g = 28$ nm) results in a sharper and higher-amplitude resonance, whereas thicker coatings (29-31 nm) broaden the peak, cause a small red shift and reduce amplitude. Therefore, an intermediate thickness of about 30 nm provides the best balance between resonance sharpness and structural stability.

After the optimization of structural parameters, the performances of designed sensor are evaluated for a range of analyte refractive indices ($n_a$ = 1.32-1.42). Fig. 5(a-b), the resonance wavelength increases with $n_a$ for both x- and y-polarizations, following excellent polynomial correlations ($R^2$ = 0.99949 for x-pol and $R^2$ = 0.99992 for y-pol). The resonance shifts from approximately 540 nm at $n_a = 1.32$ to about 1125 nm (x polarization) and 1175 nm (y polarization) at $n_a = 1.41$, extending further to 1900 nm and 2100 nm at $n_a = 1.42$. This steady red shift confirms efficient phase matching of the core and SPP modes.

The corresponding performance metrics reflect similar trends. The maximum wavelength sensitivities reach approximately 77,500 nm/RIU (x polarization) and 92,500 nm/RIU (y polarization) near $n_a = 1.41$. The amplitude sensitivities peak at around 3930 $RIU^{-1}$ and 3139 $RIU^{-1}$ for the x and y-polarizations, respectively, due to sharper loss peaks in that range. Furthermore, the figure of merit (FOM) rises sharply with increasing $n_a$, reaching 517 $RIU^{-1}$ (x-pol) and 356 $RIU^{-1}$ (y polarization).

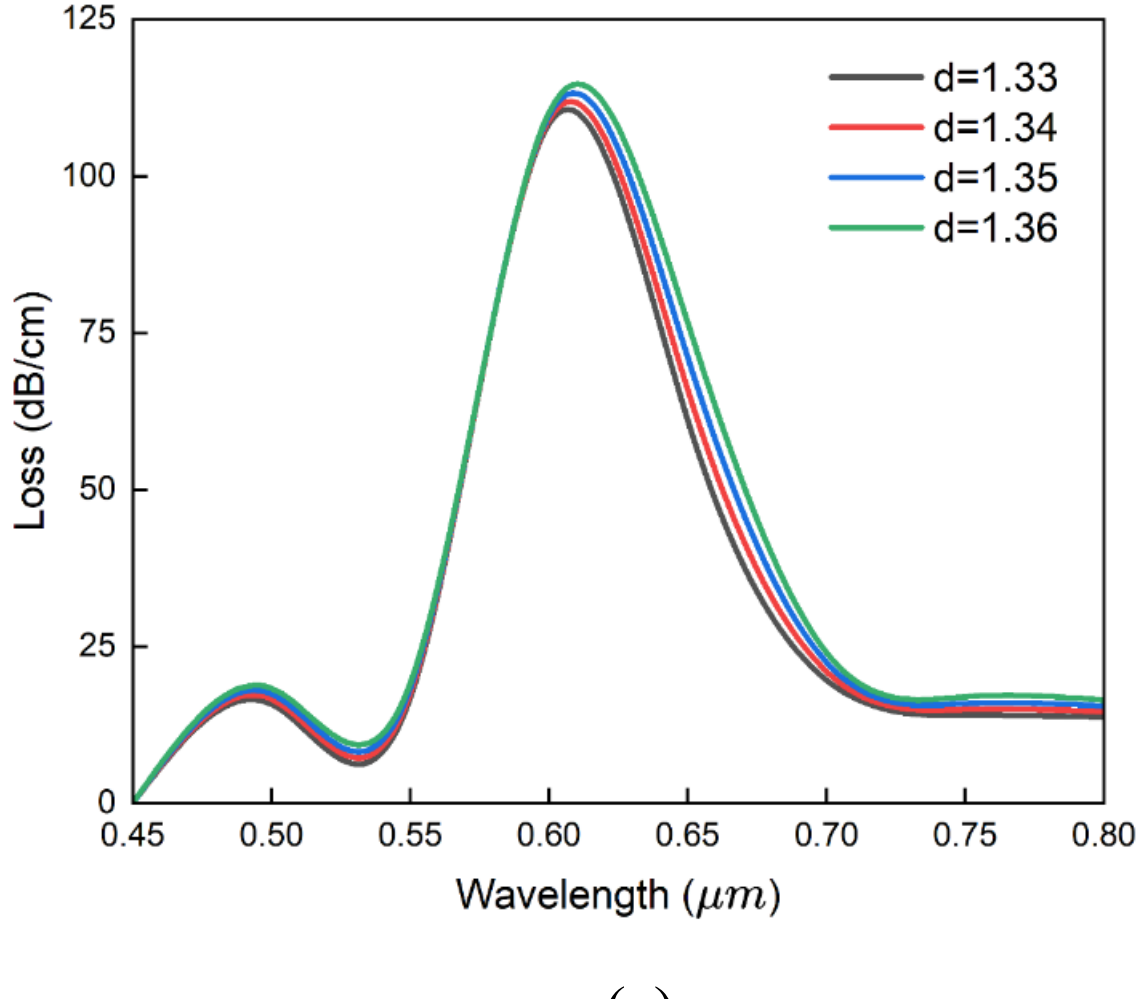

(a)

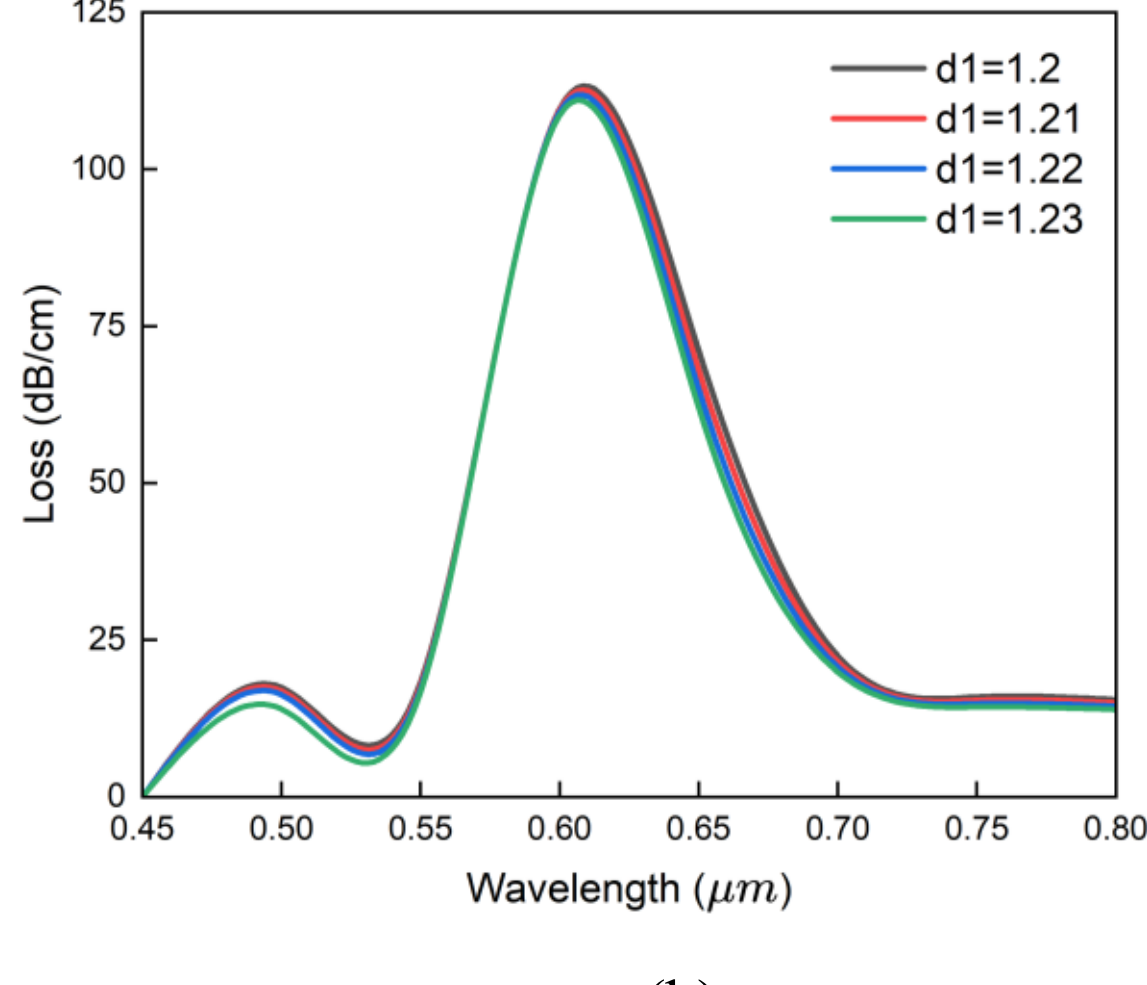

(b)

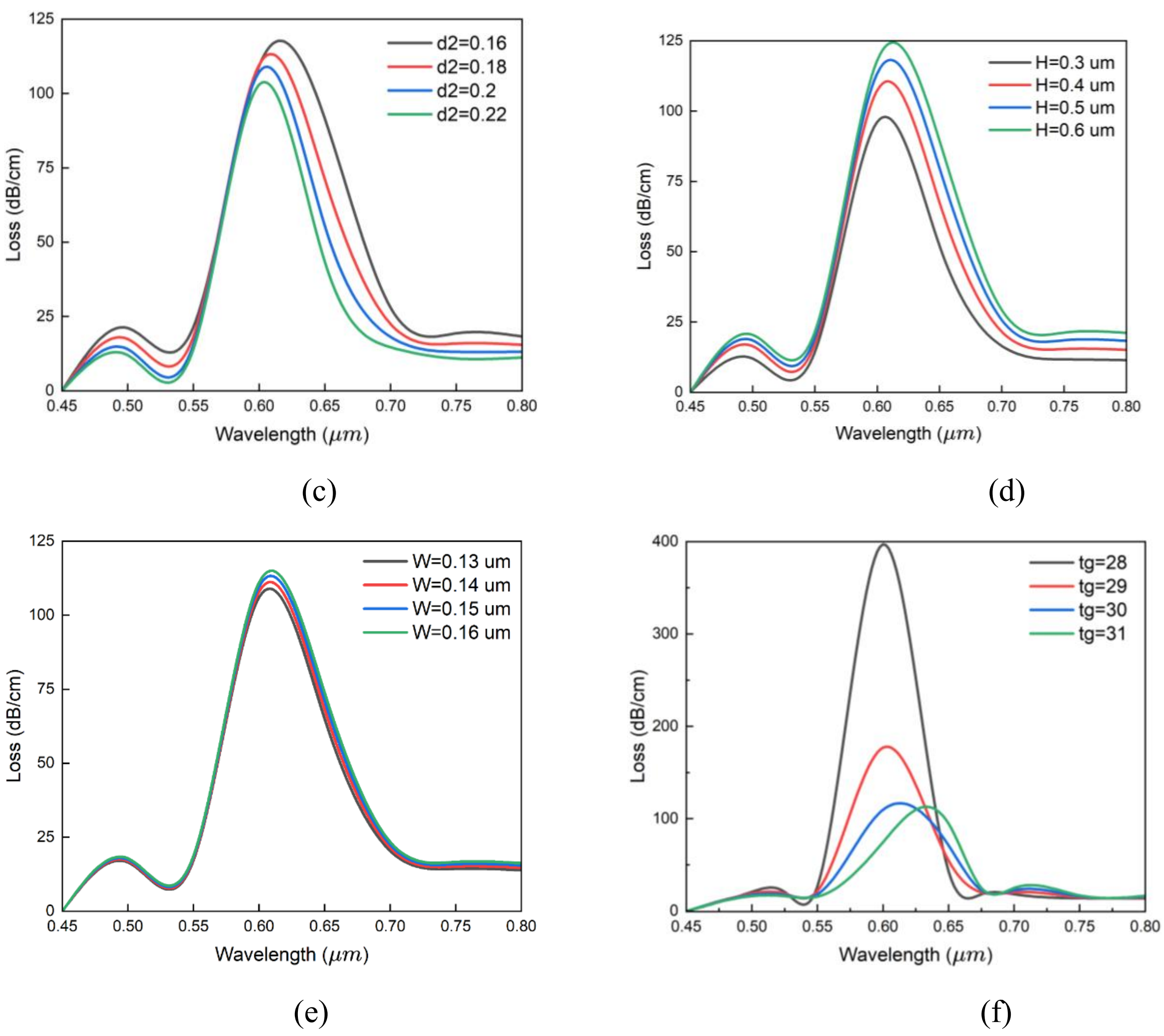

**Fig. 4:** Variation of confinement loss spectrum due to the variation of diameter of air holes (a) d (b) d2 (c) d3 (d) height and (e) width of rectangular air hole and (f) thickness of gold for $n_a$ = 1.36

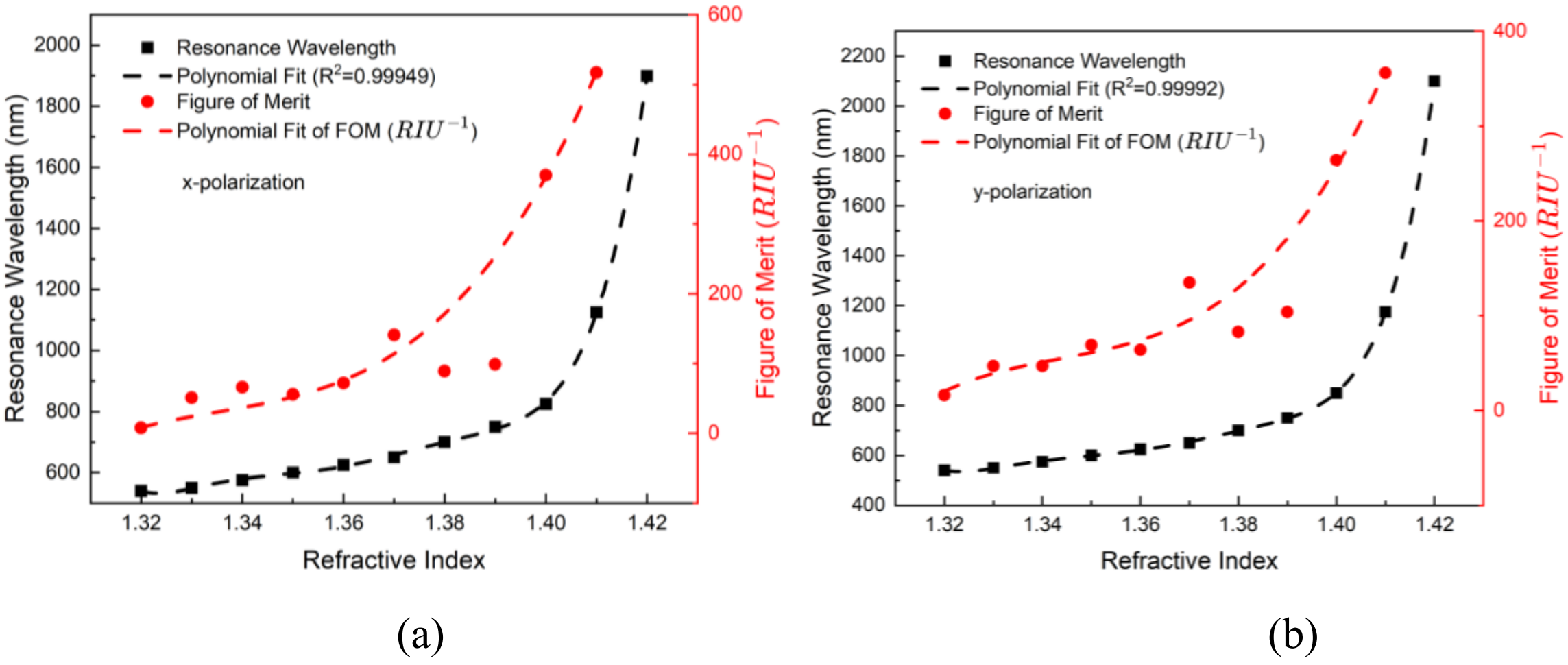

**Fig 5:** Resonance wavelength and Figure of Merit (FOM) for different values of analyte refractive indices (a) x-polarization (b) y-polarization

The proposed design is compared with several recent sensors is presented in Table 2. The result shows significantly higher sensitivity and higher FOM than previously reported designs. It shows potential for identifying cancerous cells (1.38-1.401 [48]) and in other biomedical, pharmaceutical, food and environmental sensing applications [47], [49].

**Table 2:** Comparative analysis of the proposed sensor with previously reported studies.

| Ref. | Maximum WS (nm/RIU) | Maximum AS ($RIU^{-1}$) | Maximum Resolution (RIU) | Maximum FOM ($RIU^{-1}$) | RI Range | Sensing Approach |
|---|---|---|---|---|---|---|
| [20] | 994.5 | 181.049 | $2.03 \times 10^{-6}$ | – | 1.33-1.37 | External |
| [21] | 10000 | 1115 | $2 \times 10^{-5}$ | – | 1.35-1.40 | Internal |
| [22] | 20000 | 1054 | $5 \times 10^{-6}$ | – | 1.18-1.36 | External |
| [23] | 34000 | 331 | $2.94 \times 10^{-6}$ | – | 1.16-1.37 | External |
| [24] | 53800 | 328 | $1.86 \times 10^{-6}$ | 105 | 1.14 -1.36 | External |
| [25] | 13800 | 2380 | $1 \times 10^{-6}$ | – | 1.29-1.34 | External |
| [26] | 5100 | – | $1.96 \times 10^{-5}$ | 29 | 1.19-1.40 | External |
| [17] | 62000 | 1415 | $1.61 \times 10^{-6}$ | 1140 | 1.33-1.43 | External |
| [27] | 13,000 | – | $7.69 \times 10^{-6}$ | 37 | 1.20-1.28 | External |
| [28] | 6250 | 909 | – | – | 1.36-1.392 | External |
| [50] | 24300 | 172 | $4.12 \times 10^{-6}$ | 120 | 1.10-1.45 | External |
| **This Work** | **92500** | **3930** | $\mathbf{1.08 \times 10^{-6}}$ | **517** | **1.32-1.42** | **External** |

## 5. Conclusions

This work introduces a photonic-crystal-fiber (PCF) SPR sensor that was conceived and evaluated through numerical simulation for refractive-index sensing. By integrating air holes of varying dimensions along with a rectangular slot within the core, the design facilitates stronger coupling between the guided core field and the plasmonic resonance mode. After tuning geometric parameters, the design achieves tight optical confinement, pronounced birefringence and stronger light–metal coupling. Finite-element simulations show distinct resonance features and high wavelength-sensitivity and amplitude-sensitivity responses. The sensor attains peak wavelength sensitivities of 77,500 nm/RIU (x-pol) and 92,500 nm/RIU (y-pol), along with amplitude sensitivities of 3930 $RIU^{-1}$ and 3139 $RIU^{-1}$, respectively. The corresponding figures of merit are 517 $RIU^{-1}$ and 356 $RIU^{-1}$, indicating high measurement precision and stable operation.

The design also shows strong tolerance to small fabrication changes and supports a broad refractive index (RI) sensing range as well. When compared with earlier PCF-SPR sensors, this structure provides higher sensitivity and is easier to fabricate. Moreover, its external sensing scheme simplifies analyte handling. It is noteworthy that the present work is based on simulations and idealized material assumptions. Factors such as temperature fluctuations, surface roughness or other fabrication imperfections were not considered and may affect real world performance as well. Future work will involve experimentally validating the design, adjusting it for specific biochemical sensing needs and exploring other plasmonic coating materials (like graphene, titanium nitride etc.) to further improve long term stability and extend the sensing bandwidth.